\documentclass[runningheads]{llncs}
\usepackage{booktabs} 
\usepackage{algorithm}
\usepackage{algpseudocode}
\usepackage{bm}
\usepackage{siunitx}
\usepackage{tikz}
\usepackage{adjustbox}
\usepackage[edges]{forest}
\usepackage[usenames,dvipsnames]{pstricks}
\usepackage{multirow}
\usepackage[misc]{ifsym}
\usepackage{cite}
\usepackage{amsmath,amssymb,amsfonts}
\usepackage{graphicx}
\usepackage{textcomp}
\usepackage{xcolor}
\begin{document}
\title{Unsupervised node clustering via contrastive hard sampling}

\author{Hang Cui\Letter \and Tarek Abdelzaher}

\institute{University of Illinois, Urbana Champaign\\
\email{\{hangcui2,zaher\}@illinois.edu}}

%
%

%
\maketitle              
\begin{abstract}
This paper introduces a fine-grained contrastive learning scheme for unsupervised node clustering. Previous clustering methods only focus on a small feature set (\textit{class-dependent features}), which demonstrates explicit clustering characteristics, ignoring the rest of the feature spaces (\textit{class-invariant features}). This paper exploits \textit{class-invariant features} via graph contrastive learning to discover additional high-quality features for unsupervised clustering. We formulate a novel node-level fine-grained augmentation framework for self-supervised learning, which iteratively identifies competitive contrastive samples from the whole feature spaces, in the form of \textit{positive} and \textit{negative} examples of node relations. While positive examples of node relations are usually expressed as edges in graph homophily, negative examples are implicit without a direct edge. We show, however, that simply sampling nodes beyond the local neighborhood results in less competitive negative pairs, that are less effective for contrastive learning. Inspired by counterfactual augmentation, we instead sample competitive negative node relations by creating virtual nodes that inherit (in a self-supervised fashion) \textit{class-invariant features}, while altering \textit{class-dependent features}, creating contrasting pairs that lie closer to the boundary and offering better contrast. Consequently, our experiments demonstrate significant improvements in supervised node clustering tasks on six baselines and six real-world social network datasets. 

\end{abstract}
\section{Introduction}
Node clustering, clustering graph nodes into disjoint groups, has been studied extensively in various social network applications such as polarization detection~\cite{li2022unsupervised}, anomaly detection~\cite{cui2018recursive,cui2019semi,cui2021senselens,cui2024unsupervised}, and community detection~\cite{zhang2020commdgi}. Due to the sheer volume of data and timely requirement of online applications, the problem is usually considered to be unsupervised, where no labeled nodes are available. Node clustering can also be used as graph pre-training (clustering) before labeled streams arrive. Previous node clustering methods mostly follow (1). multi-stage schemes~\cite{mrabah2023beyond,li2022unsupervised,mrabah2023contrastive,liu2022deep}, which utilize graph encoders, such as variational auto-encoders and self-supervised learning, to encode the graph unsupervised and then filter a small set of feature spaces demonstrating strong clustering characteristics, (2). integrated schemes~\cite{tsitsulin2023graph,fan2020one2multi,peng2023egrc}, which learn cluster-friendly representations via cluster-related regulators, such as modularity. However, Both schemes only focus on small feature sets, namely \textbf{class-dependent features}, while ignoring the vast \textbf{class-invariant features}, which can be further exploited to discover additional high-quality features for node clustering.

Contrastive learning has recently shown great performance on many self-supervised tasks, including image classification~\cite{chen2020simple}, graph representation learning~\cite{you2021graph}, and sequence recommendation~\cite{qiu2022contrastive}. Its core idea is to generate \textit{hard} positive and negative pairs (hard for a classifier to distinguish) that lie closer to class boundaries~\cite{wen2021toward}. By contrasting pairs that share more similarity, projection into the latent space is forced to focus more on the subset of features that truly matter in distinguishing the classes. For example, by contrasting cats with cat-sized dogs, the mapping of the respective species into the latent space is forced to consider more pertinent features to distinguish the two, than contrasting cats with, say, very large dogs, where less pertinent features (like size) may disproportionately influence the mapping. This intuition calls for fine-grained augmentations that flip the class of a node while {\em minimally changing\/} its features.

Unfortunately, previous graph contrastive learning methods~\cite{velickovic2019deep,zhu2021graph} mostly utilize coarse-grained augmentations, such as node dropping and edge perturbation. Although those methods perform well on self-supervised learning (train the graph via unsupervised contrastive learning, then freeze the trained model for downstream tasks), the obtained self-supervised representations are not clustering-friendly~\cite{zhou2022comprehensive}, requiring sufficient labeled data to fine-tune downstream tasks. 

To further obtain high-quality features for clustering, we present a novel fine-grained contrastive learning scheme that iteratively expands \textit{class-dependent features} by generating high-quality contrastive pairs. Inspired by the popular principle of conductance and modularity~\cite{tsitsulin2023graph}, which maximizes intra-cluster edges versus inter-cluster edges, we contrast each node's neighbors (as positive examples) against a selected set of negative examples, personalized to the anchor node. However, sampling negative examples is not straightforward. 
For instance, while buying a product repeatedly constitutes an endorsement, not buying it does not necessarily indicate a `dislike'~\cite {cui2021voice}. It might have other reasons, like not knowing about the product or missing the opportunity to buy it. Thus, the selection of contrasting edge absences must be done with care. 

Efficient augmentation~\cite{you2021graph} calls for generating the \textit{hardest} positive and negative pairs for the classifier/cluster to distinguish, based on the current learned parameters. Thus, the problem becomes: \textbf{what are the hardest negative examples to distinguish from positive ones?} 
Specifically, to ensure that the selected contrasting pairs are both competitive and negative, we introduce a feature decoupling scheme that captures \textit{class-invariant features} (features invariant across classes) and \textit{class-dependent features} (features variant to classes). We then create virtual nodes via augmenting \textit{class-dependent features} while keeping \textit{class-invariant features} unchanged, in order to create the most similar virtual nodes to help induce contrastive pairs. The above fine-grained augmentation method provides a parameter-free (and yet effective) scheme to generate hard positive/negative pairs.

\section{Related Work}
\label{sec:related}
\subsection{Node Clustering}
Early attempts use basic topic models, such as Bayesian models~\cite{akoglu2014quantifying} and non-negative matrix factorization~\cite{yang2020hierarchical, darwish2020unsupervised}. Later methods utilize topic modeling with cluster-specific characteristics~\cite{zhang2020commdgi} to maximize modularity (ratio of intra-cluster edges and inter-cluster edges). The recent state of the art utilizes graph neural networks~\cite{zhang2020commdgi, tsitsulin2023graph,zhu2021graph,mrabah2023contrastive,mrabah2023beyond,fan2020one2multi,liu2022deep,peng2023egrc,dou2023soft,dou2022empowering,peng2022self}, such as graph auto-encoder and contrastive learning to integrate deep neural models and cluster-specific regulators.

\subsection{Contrastive Learning}
Contrastive learning has become state-of-the-art in various self-supervised learning tasks, including image recognition~\cite{chen2020simple}, sequence recommendation~\cite{qiu2022contrastive}, and graph encoding~\cite{you2021graph}. The fundamental principle is to generate positive and negative pairs for the encoder, to help decouple desired features~\cite{wen2021toward}. The positive pairs are mostly generated as augmentations or similar neighbors. Augmentation methods aim at decoupling the correlations of spurious noise between the representations of augmented positive samples~\cite{wen2021toward}. The key challenge is to not corrupt the desired (sparse) features by augmentations. As a result, most works consult to pre-defined augmentations and distributions. 

Previous graph contrastive learning methods resort to graph corruption, such as node masking and edge perturbation, and then contrast representation learning between the original graph and corrupted graphs~\cite{zhu2021graph}. Recent works propose to use generative adversarial network (GAN) to automatically generate augmentations by creating \textit{hard} pairs for the encoder in the current epoch~\cite{you2021graph}. For domains where natural data augmentations are not available, previous literature also derive positive and negative pairs based on inferred (di)similarity to the anchor point. Examples include Mixup interpolation~\cite{verma2021towards}, contrastive predictive coding~\cite{velivckovic2018deep}, and local neighborhood~\cite{yeche2021neighborhood}.

\section{Problem Formulation and Preliminary}
\label{sec:problem}

Given an attributed graph $G(V,E,X)$ where $V$ is the vertex set, $E$ is the edge set, and $X$ is the input features of the graph. The graph is assumed to be unsigned, unweighted, and undirected since the edge weights are usually hard to acquire or estimate.

Since recent representation learning models enable easy fusion of various information, we also assume \textit{optional} auxiliary information, in the form of relational graph $G_V$ between each type of vertices. The relational graph can be constructed from (1) auxiliary relations, such as user following/friendship relation, and (2) content similarity derived from any suitable content encoders, such as BERT~\cite{peinelt2020tbert}. Given $G$ and optional $\{G_V\}$, the goal is to cluster $V$ into $K$ disjoint classes $C_k$, where $K$ is a known hyper-parameter.

Throughout this paper, we denote $\bm{H}$ as the embedding matrix, $\bm{h}$ as the embedding vector, $\bm{A} = \{a_{ij}\}$ as the adjacency matrix, $\bm{R} = \{r_{ij}\}$ as the class assignment matrix, $\mathcal{N}_u$ as the set of 1-hop neighbours of node $u$, $d_u$ as the degree of node $u$. Furthermore, we use subscript $d$ to represent matrices/vectors corresponding to \textit{class-dependent features} and subscript $o$ to \textit{class-invariant features}. 

\subsection{Graph Contrastive Learning}
Our method is based on self-supervised graph contrastive learning\cite{zhu2021graph,you2021graph}, which is a learning scheme consisting of two steps: \\
(1) Contrastive self-supervised learning: The graph is encoded unsupervised on a contrastive loss. The most popular graph encoding method is Graph Convolutional Netowrk(GCN)~\cite{hamilton2017inductive}

For an anchor node $\bm{v}$, the objective of contrastive learning is:
\begin{align}
    \mathcal{L}_{CE} = \mathbb{E}_{u^+\sim P_p(u|v),u^-_k\sim P_n(u|v)} [-\log\frac{e^{f(\bm{v})^\top f(\bm{u}^+)/\tau}}{\sum_k e^{f(\bm{v})^\top f(\bm{u}^-_k)/\tau}}]
    \label{eq:ns}
\end{align}
where $\tau$ is a temperature parameter, $P_p$ and $P_n$ are (often pre-defined) positive and negative distributions, respectively.\\

\section{MeCole}
\label{sec:motive} 
Previous node clustering methods~\cite{zhang2020commdgi, tsitsulin2023graph,zhu2021graph,mrabah2023contrastive,mrabah2023beyond,fan2020one2multi,liu2022deep,peng2023egrc} focus on a small set of features demonstrating strong clustering characteristics, such as modularity and conductivity; while ignoring the rest of the feature spaces. Our method proposes a feature expansion scheme that iteratively \textbf{expands} the set of features for clustering: 
\begin{itemize}
    \item Start with a small set of \textbf{high-confident features}. The high-confident feature sets include two sub-components (1). \textbf{class-dependent feature set}: features of strong clustering characteristics (2). \textbf{class-invariant feature set}: features of minimal clustering characteristics. 
    \item Expand the high-confident feature set on both components. In other word, our method iteratively discovers additional \textit{class-dependent features} and \textit{class-invariant features} from the rest of feature spaces.
\end{itemize} 

\subsection{Node-level fine-grained contrastive learning}
We propose a novel node-level fine-grained augmentation, which can efficiently generate \textit{hard} positive/negative examples for any given anchor nodes. Our objective is to contrast between positive and negative examples of node relations, which are in line with the maximization of modularity and conductance (ratio of intra-class edges against inter-class edges). The positive node relations are sampled via edge connection. The negative node relations, however, are considerably more sophisticated to understand, due to the unavailability of associated content and the multitude of potential reasons for not having an edge, such as lack of observations and attention. Therefore, the core contribution of this paper lies in how to \textbf{generate a selected set of negative node relations that are competitive for contrastive learning}.

Our method is inspired by \textbf{counterfactual augmentations}~\cite{ma2022learning, mesnard2020counterfactual}: how would the node interactions change if any feature(s) is augmented? Given an anchor node $v$, we synthesize a set of virtual nodes $\{\bar{v}\}$ via feature augmentation(s) as counterfactual examples of $v$. The deviation of node interaction patterns between the anchor node and the virtual nodes represents the counterfactual interactions, which are exploited to sample negative node relations.

To ensure the virtual nodes satisfy both negative and competitive, we propose to synthesize nodes via \textbf{minimal augmentation}: highly similar to $v$ (competitiveness), but are clustered into other classes or the boundary (negativity) under the current learned parameters. We summarize the above idea into 4 conditions to generate virtual nodes $\bar{v}$:
\begin{itemize}
    \item Competitiveness: maximize similarity to the anchor node $v$. There are two popular options for such similarity metrics: (1). L1 norm: the sum of absolute augmentations (2). L0 norm: the number of augmented features.
    \item Negativity: minimize the likelihood of being clustered to the same class as $v$. The reason is to ensure the synthesized node is true negative so that the sampled negative relations are not false negative.
    \item Randomness: minimize mutual information between pairs of synthesized virtual nodes. This is a contrastive learning condition to improve efficiency.
    \item Parameter-free: the generator model has few or no additional parameters. This is to reduce over-fitting in the self-supervised setting.
\end{itemize}
While the above conditions can be satisfied in various ways with different trade-offs, we provide a parameter-free yet effective scheme in this paper as a foundation of our framework.  

\subsection{Augmentation Scheme}
The augmentation scheme is based on the \textbf{high-confident feature sets}: \textit{class-invariant features} and \textit{class-dependent features}. Given an anchor node, virtual nodes are created by randomly augmenting \textit{class-dependent features} to randomly sampled nodes clustered into the opposing classes; while keeping \textit{class-invariant features}. The intuition is: (1). \textit{class-dependent features} serve as the short paths towards the opposing classes, which enables \textit{Negativity} while minimizing features augmented to satisfy \textit{Competitiveness}. (2) \textit{class-invariant features} facilitate the learning of high-quality \textit{class-dependent features} in the joint training of contrastive objective. (3). The random sampling scheme ensures \textit{Randomness} condition. (4). The only parameter is the sampling probability, which can be dynamically adjusted on the go.

We then contrast the positive examples of node relations (what the anchor node interacts with) with the negative examples of node relations: nodes that the anchor node does not interact with but the virtual node would interact with.

\subsection{Model Overview}

\begin{figure*}
\includegraphics[width=\textwidth,trim=1.5cm 7cm 9cm 3cm,clip]{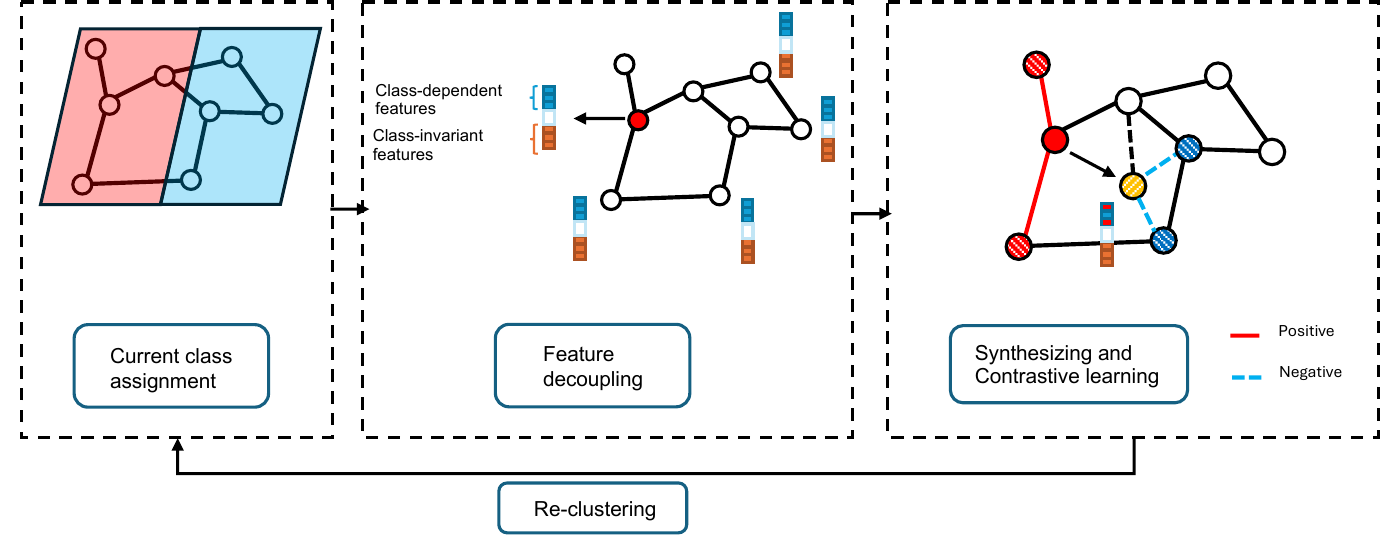}
\caption{MeCoLe: (1). Class assignment of last epoch (2). Learn decoupled \textit{class-invariant features} and \textit{class-dependent features} from the underlying graph, contrastive learning loss, and class assignments. (3). Randomly augment (from anchor node \textcolor{red}{red}) \textit{class-dependent features} towards nodes in opposing classes to generate synthesized virtual nodes (\textcolor{yellow}{yellow}-shaded) and generate edges for the synthesized nodes. Derive contrastive learning loss between positive pairs (\textcolor{red}{red}-shaded) and negative pairs (\textcolor{blue}{blue}-shaded).}
\label{MeCoLe}
\end{figure*}

Our framework consists of three integrated modules: feature decoupling module, synthesizing and contrastive learning module, and decoupled cluster module. Feature decoupling module decouples \textit{class-dependent features} (features correlate with the downstream task) and \textit{class-invariant features} (features that do not correlate). Synthesizing and contrastive learning module generates a set of virtual nodes per anchor node and then samples a set of negative nodes (nodes a virtual node would interact with but the anchor node does not) as negative relations and then contrasts between the sampled negative nodes and sampled positive nodes (nodes the anchor node interacts with). The decoupled cluster module filters out irrelevant nodes and updates node assignments into the $K$ classes.

In principle, we can use any previous clustering techniques to initialize the clusters. By default, we use GCN with modularity regulator~\cite{tsitsulin2023graph}. In each iteration:
\begin{itemize}
    \item The feature decoupling module updates learned decoupled features based on current class assignments.
    \item The synthesizing and contrastive learning module synthesizes virtual nodes, and then samples fine-grained positive/negative pairs for contrastive learning.
    \item The decoupled cluster module updates class assignments.
\end{itemize} 

\subsection{Feature Decoupling}
\label{sec:decouple}
Given the graph inputs $\{G\}$, our graph encoding module deploys a decoupling framework to output \textit{class-dependent features} and \textit{class-invariant features} according to the current class assignments. 

We design the following generalized objective to decouple learned features:
\begin{subequations}
\label{objective}
\begin{align}
    &\bm{H}_d,\bm{H}_o\sim {enc}(G)
    \label{eq:Md}\\
    &min_{v_1,v_2\sim V}\mathcal{L}_1(G, Z(\bm{H}_d,\bm{H}_o))
    \label{eq:L}\\
    &min\mathcal{L}_2 = min_{v_1\sim C_1,v_2\sim C_2}  \frac{d(\bm{h}_{o,1},\bm{h}_{o,2})}{d(\bm{h}_{d,1},\bm{h}_{d,2})}
    \label{eq:D}
\end{align}
\end{subequations}
where $\bm{H}_d$ and $\bm{H}_o$ are the matrix of \textit{class-dependent features} and \textit{class-invariant features}, respectively; ${enc}$ is the graph encoder; $C_k$ are the vertex set of class $k$; $Z$ is a link prediction model that reconstructs $G$ from $\bm{H}_d$ and $\bm{H}_o$; $\mathcal{L}_1$ is the reconstruction loss, such as cross entropy; $d$ is a distance metric that measures discrepancy between nodes in two classes. The above formulation decouples the self-supervised models on class-invariant and class-dependent signals, enabling easy adaptation of existing state-of-the-art graph models.

The formulation is solved as a joint training framework between $\bm{H}_d,\bm{H}_o$. Eq.(\ref{eq:L}) minimizes the reconstruction loss of $G$ and $Eq.(\ref{eq:D})$ disentangles \textit{class-dependent features} and \textit{class-invariant features} by minimizing the relative discrepancies. We use two separate GCNs as our default graph encoder on both feature sets.

\subsubsection{Link prediction model}
\label{sec:prediction}
Graph reconstruction is commonly used in clustering techniques to transfer knowledge from connection patterns~\cite{pan2023beyond}. The link prediction model $Z$ reconstructs graph $G$ by estimating the probability of edges between node pairs. The main question is how to combine the \textit{class-dependent features} $\bm{H}_d$ and \textit{class-invariant features} $\bm{H}_o$ in the prediction. Our experiments show that a simple black box classifier does not perform well due to the lack of labeled data in an unsupervised setting.

We observe that: in online scenarios, two nodes interact when both \textit{class-dependent features} and \textit{class-invariant features} are satisfied. For example, a Republican supporter supports online items when the post satisfies both \textit{class-dependent features} (i.e., the post favors Republican) and \textit{class-invariant features} (i.e., the post engages him;  the post is within the topic of interest, etc.). Therefore, we use a simple yet effective prediction model $Z$:
\begin{align}
    Z_{uv}(\bm{H}_d,\bm{H}_o) \sim Z_{uv}^d(\bm{H}_d)\times Z_{uv}^o(\bm{H}_o)
    \label{eq:prediction}
\end{align}
The above formulation further disentangles \textit{class-dependent features} and \textit{class-invariant features} in probability relations between both sets of features, while also enabling easy adaptation of state-of-the-art predictors. We use dot product for $Z_{uv}^d,Z_{uv}^o$, because of its state-of-the-art performance on self-supervised tasks.

\subsubsection{Discrepancy metric}
\label{sec:discrepancy}
The most popular discrepancy metrics are $l$-norm and cosine similarity. Examples are: $1$-norm (commonly used by domain adaptation); euclidean distance $2$-norm; and max-pooling $\infty$-norm. Our experiment shows our method is resilient to the above choices. Nevertheless, the best performance comes from $1$-norm.

\subsection{Joint Learning Framework}
\label{sec:joint}
Our decoupled representation learning framework jointly optimizes the objectives eq.(\ref{objective}) in a self-supervised manner. Following the discussion in section.~\ref{sec:prediction}, we follow a graph rewiring procedure~\cite{bruel2022rewiring}. In short, given the input graph $G$ and the current estimation of $\bm{H}_o$. We rewire $G$ by $G^d = f(G|G^o)$, where $G^o$ is the reconstructed graph from $\bm{H}_o$; and then use $G^d$ to update $\bm{H}_d$ in the iteration. From eq.(\ref{eq:prediction}), the back-propagation is:
\begin{align}
    \frac{\delta Z}{\delta \bm{H}_d} = Z^o(\bm{H}_o)\frac{\delta Z^d}{\delta \bm{H}_d}
\end{align}
Therefore, the augmentation function $f$ is:
\begin{align}
    e^d_{ij} = f(e_{ij}|e^o_{ij}) = min(\eta, e_{ij}/e^o_{ij})&  &\forall e_{ij}\in G
\end{align}
where $e^o_{ij}=Z^o(\bm{h}^o_{v_i},\bm{h}^o_{v_j})$ and $\eta$ is the maximum weight of an edge in the rewired graph.

\subsection{Integrate Content Representations}
\label{sec:content}
Content latent representations can facilitate the decoupling of \textit{class-dependent features} and \textit{class-invariant features}. We assume the contents are encoded into: (1). latent representations $X$ using suitable content encoders, such as BERT (2). a sequence of key attributes $M$.

Due to the limited fine-tuning inputs in self-supervised settings, we only focus on extracting high-quality features. For latent representation $X$, we first construct the content similarity graph to enrich the input graphs:
\begin{align}
    \bm{A}_{ij}^{sim} \sim sim(\bm{X}_i,\bm{X}_j)
\end{align}

To filter out low-quality features, we first construct the $k$-nn graph, such that we only keep the top-$k$ edges of a node. The graph is further filtered by a threshold $\eta_{1}$ by removing any edges of similarity lower than $\eta_{sim}$. The graph $G_{X}$ is then included into the heterogeneous graph encoder in section.~\ref{sec:decouple}.

The key attributes $M$ are separated according to its distribution. We use \textit{tf-idf} to separate key attributes. In details, we compute \textit{tf-idf} scores $\bm{S}_k$ of attributes for each classes $k$ and then derive the class-dependent representation of node $i$ as:
\begin{align}
    \bm{M}_{d,i} = \sum_k r_{ik} \frac{\sum_{m\in M} S_{k,m}\bm{x}_m}{\sum_{m\in M}S_{k,m}}
\end{align}
where $x_m$ is the word embedding of attribute $m$ and $r_{ik}$ is the class assignment of node $i$ to class $k$. The key attribute \textit{class-dependent features} are then constructed into $G_M$ in the same fashion as $G_X$.

\subsection{Synthesizing nodes and Contrastive Learning}
Anchor node selection favors the \textit{boundary} nodes. In detail, for each class, we perform Gaussian sampling from nodes within the class according to the inverse assignment to the class. Given the anchor node $v$, we generate positive samples from its 1-hop neighbors. The reason is to demonstrate that the most simple and generalizable positive sampling scheme is sufficient to produce good results. For negative samples, we first create virtual nodes $\tilde{v}$ by randomly augmenting \textit{class-dependent features} with probability $P_{CE}$ to uniformly sampled nodes in the other classes
then, sample negative nodes proportional to the edge prediction probability $Z(\tilde{v},u)$.
\begin{align}
    &\mathcal{L}_{CE} = \mathbb{E}_{u^+\sim P_p(u|v),u^-_k\sim P_n(u|v)} [-\log\frac{e^{f(\bm{v})^\top f(\bm{u}^+)/\tau}}{\sum_k e^{f(\bm{v})^\top f(\bm{u}^-_k)/\tau}}]\\
    &P_p(u^+|v) = 
    \begin{cases}
      1/d_v & u^+\in \mathcal{N}_v\\
      0 & u^+\notin \mathcal{N}_v
    \end{cases} 
    \\
    &P_n(u^-|v) \sim Z(\tilde{v},u) \quad u^-\in \mathcal{N}_{\tilde{v}}-\mathcal{N}_v
\end{align}
where $f(v) = \bm{h}_{d,v}$ is the class-dependent feature of node $v$, $\mathcal{N}_v$ is the set of neighboring nodes of $v$, $P_p$ and $P_n$ are the positive and negative distributions respectively. The implication is to distinguish the nodes that $v$ interact with and the nodes the virtual nodes $\tilde{v}$ interact with.

\subsection{Decoupled Cluster Module}
\label{sec:cluster}
The cluster module is based on the previous contrastive learning framework~\cite{zhu2021graph}: logistic regression for each class.

Compared to previous works~\cite{tsitsulin2023graph,fan2020one2multi,peng2023egrc}, we do not include any regulators, such as modularity and conductance, to encourage clustering-friendly representations. The reason is two-folded:
\begin{enumerate}
    \item Our contrastive objective already includes the modularity principle by contrasting positive and negative node relations.
    \item Static-form regulators often interfere the self-supervised objectives. 
\end{enumerate}

\subsection{Put Everything Together}
To sum up, the loss function consists of three components:
\begin{enumerate}
    \item Graph loss: $\mathcal{L}_G = \mathcal{L}_1+\mathcal{L}_2$
    \item Contrastive learning loss: $\mathcal{L}_{CE}$
\end{enumerate}
Our model minimizes:
\begin{align}
    \mathcal{L} = \mathcal{L}_G + \alpha_{CE}\mathcal{L}_{CE}
\end{align}

\section{Experiments}
\label{sec:exp}
This section evaluates our framework on four public social network datasets \textbf{Wikipedia}\footnote{http://konect.cc/networks/}, \textbf{IMDB}\footnotemark[\value{footnote}], \textbf{Covid}~\cite{cui2021voice}, \textbf{Protest}~\cite{cui2021voice}, two citation datasets \textbf{Cora}, \textbf{Citeseer}, and two newly collected datasets. The two new datasets differ from the previous ones, by including a substantial number of irrelevant nodes. The purpose is to simulate the collected raw datasets when the understanding of classes is minimal.
\begin{itemize}
    \item \textbf{2020 presidential election}: is a newly collected data set from Twitter API using keywords: `presidential election', `Trump', and `Biden'. Nodes are users and tweets, and an edge denotes that the user originally posts or positively interacts with the tweets.
    \item \textbf{Police shooting}: is a newly collected data set from Twitter API on the shooting of Daunte Wright, using keywords: `police shooting', `Daunte Wright'. Nodes are users and tweets, and an edge denotes that the user originally posts or positively interacts with the tweets.
\end{itemize}

\begin{table}[]
\caption {\% Dataset statistics}
\centering
\label{tab:stats}
\begin{tabular}{|l|l|l|l|l|l|l|l|l|}
\hline
                   &\text{\#nodes}    &\#edges &\#classes &Ave. deg &Inputs \\ \hline
Wiki        & \text{1124}      & \text{6205}    & \text{3} & \text{11.04} &  $G,G_X$\\ \hline
IMDB      & \text{12875}      & \text{40711}    & \text{6} & \text{6.81} & $G,G_X$ \\ \hline
Covid         & \text{1175}      & \text{4429}    & \text{3} & \text{7.53} &$G,G_V,G_X$ \\ \hline
Protest & \text{1025}      & \text{2245}    & \text{3} & \text{4.38} &$G,G_V,G_X$ \\ \hline
President  & \text{2063}     & \text{11883}    & \text{3} & \text{11.52} &$G,G_V,G_X$ \\ \hline
Police  & \text{2011}      & \text{12438}    & \text{4} & \text{12.37} &$G,G_V,G_X$ \\ \hline
Cora & \text{2485}      & \text{5069}    & \text{7} & \text{4.07} &$G$ \\ \hline
Citeseer & \text{2110}      & \text{3668}    & \text{6} & \text{3.48} &$G$ \\ \hline
\end{tabular}
\end{table}

In addition to graph $G$, we also include auxiliary graphs constructed from node relations ($G_V$) and contents ($G_X$) using the methods described in section.~\ref{sec:content}. The detailed statistics of datasets are shown in Table.~\ref{tab:stats}.

We compare our method with unsupervised clustering methods: \textbf{DMoN}~\cite{tsitsulin2023graph}, \textbf{DCRN}~\cite{liu2022deep},
\textbf{DGCN}~\cite{pan2023beyond},
\textbf{CommDGI}~\cite{zhang2020commdgi}, \textbf{vGAE}~\cite{li2022unsupervised}, \textbf{cvGAE}~\cite{mrabah2023contrastive}; and self-supervised methods: \textbf{CCA-SSG}~\cite{zhang2021canonical}, \textbf{GRACE}~\cite{zhu2021graph}, \textbf{InfoGCL}~\cite{xu2021infogcl}. \\

\subsection{Experiment Results}
\label{sec:main_result}

\begin{table}[]
\caption {Clustering accuracy}
\centering
\label{table:exp}
\begin{tabular*}{\linewidth}{@{\extracolsep{\fill}} l|l|l|l|l|l|l|l|l}
\hline
                   &Wiki    &IMDB &Covid & Protest &President &Police & Cora & Citeseer \\ \hline
DMoN          & 69.5      & 75.6    & 70.1 & 69.6 & 75.9 & 75.0 & 46.3 & 69.5 \\ \hline
DCRN          & 71.9      & 76.5    & 73.0 & 70.5 & 75.6 & 74.9 & 48.3 & 70.9 \\ \hline
DGCN          & 74.2      & 77.6    & 76.2 & 70.5 & 79.1 & 75.2 & 72.9 & 71.6 \\ \hline
CommDGI          & 78.5      & 80.3    & 80.1 & 72.1 & 82.9 & 76.2 & 70.1 & 69.4 \\ \hline
vGAE   & 74.0      & 78.1    & 76.5 & 70.6 & 78.6 & 75.1 & 73.1 & 68.2 \\ \hline
cvGAE    & 76.3      & 80.1    & 77.4 & 72.1 & 80.6 & 76.2 & 79.0 & 71.8 \\ \hline
CCA-SSG    & 79.5      & 80.3    & 76.2 & 71.0 & 77.6 & 79.1 & 70.5 & 68.7 \\ \hline
GRACE    & 74.2      & 76.1    & 73.4 & 69.5 & 77.4 & 75.4 & 65.8 & 67.5 \\ \hline
InfoGCL    & 78.9      & 80.1    & 75.5 & 70.4 & 77.5 & 78.0 & 69.6 & 69.1 \\ \hline
MeCoLe    & $\bm{86.8}$      & $\bm{89.7}$   & $\bm{84.3}$ & $\bm{80.1}$ & $\bm{90.7}$ & $\bm{83.7}$  & $\bm{79.9}$ & $\bm{74.6}$ \\ \hline
\end{tabular*}
\end{table}

The results are presented in Table.~\ref{table:exp}. Our approach outperforms 9 baselines in all 8 datasets. The contrastive methods, CCA-SSG, GRACE, and InfoGCL, outperform unsupervised clustering methods in the 6 social network datasets, while underperform in the 2 citation datasets. The observation demonstrates their strong capability of extracting robust features from noisy real-world datasets. Our method (1). integrates both contrastive objectives and traditional clustering objectives (2). utilizes the vast \textit{invariant features} to facilitate the learning of \textit{dependent features}; and thus outperform other baselines.

Out of the 6 social network datasets, \textit{Protest} has the lowest clustering accuracy because of the lowest average degree of the underlying graph. \textit{Police} also shows relatively low accuracy despite the high average degree. The reason is the intertwined opinions on related topics across classes. 

\subsection{Ablation Study}

\begin{table}[]
\caption {Clustering accuracy - ablation study}
\label{table:abl}
\begin{tabular*}{\linewidth}{@{\extracolsep{\fill}} l|l|l|l|l|l|l|l|l}
\hline
                   &Wiki    &IMDB &Covid & Protest &Pres. &Police & Cora & Citeseer \\ \hline
Baseline    & 86.8      & 89.7    & 84.3 & 80.1 & 90.7 & 83.7 & 79.9 & 74.6\\ \hline
Feature decouple          & 74.1      & 77.1    & 77.2 & 72.1 & 83.3 & 75.2 & 70.2 & 66.0 \\ \hline
Negative sample      & 73.2      & 77.2    & 77.6 & 76.2 & 82.1 & 74.4 & 72.5 & 66.4\\ \hline
MLP          & 82.2      & 87.1    & 80.9 & 83.7 & 85.5 & 78.6 & 76.3 & 72.6 \\ 
\hline
- C.L.           & 75.0      & 76.9    & 76.8 & 71.3 & 79.2 & 74.3 & 70.1 & 69.0\\ 
\hline
Graph augment        & 81.9      & 83.1    & 78.5 & 72.4 & 80.5 & 80.0 & 75.6 & 71.2 \\ \hline
-$G_V$   & \text{-}      & 86.4    & 81.1 & 74.7 & 86.2 & 78.1 & \text{-} & \text{-} \\ \hline
-$G_X$    & 84.2      & 88.1    & 81.5 & 77.4 & 87.9 & 80.1 & \text{-} & \text{-} \\ \hline
\end{tabular*}
\end{table}

This section performs the ablation analysis of each module, including:
\begin{itemize}
    \item \textbf{Feature decoupling} is a core module of our framework. We remove the decoupled representation learning and (1) augment all learned features in contrastive learning (2). modifies negative sampling approaches to sample any nodes outside of neighborhood (3) replace the prediction model with MLP.
    \item \textbf{Contrastive Learning}. We first (1) completely remove contrastive learning (the clustering uses the output of the graph encoder), and then (2) replace our node-level augmentations with graph augmentation approaches.
    \item \textbf{Auxiliary graph}. We remove (1) user-user relations $G_s$ (2) item-item relations $G_t$ (3) content representation $X$ (4) key attributes $W$.
\end{itemize}

The results are presented in Table.~\ref{table:abl}. In general, our approach can perform relatively well without any of the above components. Out of all modules, the feature decoupling module shows the most significant performance drops, which supports our claim that \textit{hard} positive/negative pairs are essential for self-supervised contrastive learning. Compared to graph-level augmentations~\cite{zhu2021graph}, our methods are superior in: (1) generating effective contrastive learning pairs of any anchor nodes, (2) utilizing the downstream clustering results to improve pairs generation. Our methods are also resilient in removing input types. The experiment shows $G_V$, the user relation graph is relatively more important to the performance. The reason behind this is the lack of content information on users (i.e., user profiles) in the datasets.

\subsection{Discrepancy Functions}
This section evaluates the discrepancy function $d$ used in our method. We focus on $l$-norm and \textit{cosine} in this paper because of their simple yet state-of-the-art performance in machine learning models.

\begin{table}[]
\caption {Clustering accuracy - discrepancy function}
\label{table:dis}
\begin{tabular*}{\linewidth}{@{\extracolsep{\fill}} l|l|l|l|l|l|l|l|l}
\hline
                   &Wiki    &IMDB &Covid & Protest &Pres. &Police & Cora & Citeseer \\ \hline
Baseline ($l=1$)    & 86.8      & 89.7   & 84.3 & 80.1 & 90.7 & 83.7 & 79.9 & 74.6 \\ \hline
$l=2$         & 84.3      & 86.8    & 80.7 & 76.9 & 84.3 & 79.2 &79.8 & 74.8 \\ \hline
cosine    & 83.5      & 86.1    & 79.5 & 74.9 & 82.2 & 81.4 & 79.5 & 74.2 \\ \hline
$l=\infty$     & 81.6      & 82.2    & 71.1 & 70.0 & 79.8 & 77.2 & 78.2 & 74.2\\ \hline
\end{tabular*}
\end{table}

Our results in table.~\ref{table:dis} demonstrates $l=1$-norm being the best discrepancy for feature decoupling. The result is in line with previous literature on feature-based domain adaptations~\cite{ganin2015unsupervised}.

\subsection{Integrate Contrastive Learning}
Our contrastive learning framework can be integrated into most graph encoders. In detail, we replace $enc$ in eq.~(\ref{objective}) with the baseline encoders and use the joint learning method described in section.~\ref{sec:joint} to train the encoders jointly. For GNN-based methods (COMMDGI, GCN, and vGAE), this is a straightforward adaptation.

\begin{table}[]
\caption {Clustering accuracy - integrating contrastive learning}
\label{table:IT}
\begin{tabular*}{\linewidth}{@{\extracolsep{\fill}} l|l|l|l|l|l|l|l|l|l}
\hline
                 &  &Wiki   &IMDB & Covid & Protest &Pres. &Police & Cora & Citeseer\\ \hline
\multirow{ 2}{*}{C.DGI}          &   & 78.5      & 80.3    & 80.1 & 72.1 & 82.9 & 76.2 &70.1 &69.4\\& +CL
& 84.4      & 85.3   & 83.2 & 79.1 & 88.0 & 80.2 &71.5 &70.9\\\hline
\multirow{ 2}{*}{GCN}          &   & 66.0      & 68.9    & 70.8 & 64.3 & 71.2 & 65.3 &44.2 & 51.3 \\& +CL
& 75.2      & 78.6    & 78.9 & 74.2 & 83.9 & 74.3 &60.2 & 67.2\\\hline
\multirow{ 2}{*}{vGAE}          &   & 74.0      & 78.1    & 76.5 & 70.6 & 78.6 & 75.1 &73.1 &68.2\\& +CL
& 79.3      & 84.2    & 80.7 & 76.7 & 85.1 & 78.3 & 76.6 & 70.9\\\hline
MeCoLe          &   & \textbf{86.8}      & \textbf{89.7}    & \textbf{84.3} & \textbf{80.1} & \textbf{90.7} & \textbf{83.7} & \textbf{79.9} & \textbf{74.6}\\\hline
\end{tabular*}
\end{table}

The results are presented in Table.~\ref{table:IT}. In general, our node-level contrastive learning method demonstrates performance boosts on all 3 baselines on 8 datasets. GNN-based methods benefit from contrastive learning because of their robust representation models on integrated inputs. The sparest dataset, \textit{Protest}, shows the highest performance boosts on all baselines, demonstrating the benefits of contrastive learning on sparse graphs. Another observation is that GCN enjoys a greater performance boost compared to other baselines. Without contrastive loss or cluster-specific regulations, GCN fails to learn robust cluster-friendly representations.

\subsection{Sparse Graph}
This section evaluates the clustering performance on the sparse section of the datasets. In detail, we remove the top $30\%$ degreed nodes from the datasets and self-training, then evaluate the clustering accuracy of the rest of the datasets. We exclude citation datasets, Cora and Citeseer, due to the numerous disconnected components.

\begin{table}[]
\caption {Clustering accuracy - Sparse}
\label{table:SP}
\begin{tabular*}{\linewidth}{@{\extracolsep{\fill}} l|l|l|l|l|l|l}
\hline
                   &Wiki    &IMDB &Covid & Protest &President &Police \\ \hline
DMoN     & 55.4      & 62.3    & 65.5 & 68.8 & 70.8 & 72.2 \\ \hline
DCRN     & 58.2      & 64.1    & 68.0 & 67.8 & 70.9 & 72.5 \\ \hline
DGCN     & 67.2      & 72.5    & 70.5 & 69.5 & 71.2 & 71.9 \\ \hline
CommDGI     & 70.9      & 76.3    & 76.5 & 69.5 & 78.6 & 72.8 \\ \hline
vGAE   & 68.0      & 74.2    & 73.5 & 67.4 & 75.9 & 70.0 \\ \hline
cvGAE    & 70.1      & 73.6    & 74.2 & \text{0.702} & 78.5 & 70.8 \\ \hline
CCA-SSG    & 74.5      & 80.5    & 76.2 & 70.7 & 78.7 & 76.7 \\ \hline
GRACE  & 72.2      & 77.9    & 74.8 & 68.9 & 77.1 & 76.3 \\ \hline
InfoGCL    & 71.6      & 75.3    & 71.2 & 66.7 & 76.7 & 74.7 \\ \hline
MeCoLe    & $\bm{81.8}$      & $\bm{87.6}$   & $\bm{82.9}$ & $\bm{79.2}$ & $\bm{86.7}$ & $\bm{80.3}$ \\ \hline
\end{tabular*}
\end{table}

The results are presented in table.~\ref{table:SP}. Our method shows greater resilience on sparse graphs than other baselines. The unsupervised clustering methods, DMoN, DCRN, and DGCN, suffer greater performance loss in sparse graphs due to their heavy reliance on message-passing models to extract cluster-specific features.

\section{Conclusion}\
\label{sec:conclusion}
This paper presented MeCoLe, a novel fine-grained contrastive learning framework that integrates the commonly ignored \textit{class-invariant features} with \textit{class-dependent features} in unsupervised node clustering tasks. MeCoLe innovates as the first attempt to formulate a finer-grained contrastive learning scheme on node relations: contrast between edges (as positive node relations) and absent edges (no edge as negative relations) as a source of contrastive learning to improve node clustering tasks. To generate competitive contrastive pairs, MeCoLe first decouples \textit{class-dependent features} and \textit{class-invariant features}, then generates virtual nodes by augmenting \textit{class-dependent features} while retaining \textit{class-invariant features}, and finally sample the negative node relations as counterfactual relations.

\subsubsection*{Acknowledgements.} Research reported in this paper was sponsored in part by DARPA award HR001121C0165, DARPA award HR00112290105, and DoD Basic Research Office award HQ00342110002. It was also supported in part by ACE, one of the seven centers in JUMP 2.0, a Semiconductor Research Corporation (SRC) program sponsored by DARPA
%
%
%
%
\bibliographystyle{IEEEtran}
\bibliography{main}
\end{document}